\documentclass[conference,final]{IEEEtran}

\usepackage{booktabs} %
\usepackage{listings}
\usepackage{fancyvrb}
\usepackage{wrapfig}
\usepackage{graphicx}
\usepackage{amsmath}
\usepackage{amssymb}
\usepackage{color}
\usepackage{ifpdf}
\usepackage{subcaption}
\usepackage{float}
\usepackage[utf8]{inputenc}
\usepackage{multirow}
\usepackage{rotating}

\usepackage{xcolor}
\PassOptionsToPackage{hyphens}{url}
\usepackage[colorlinks,breaklinks=true]{hyperref}
\AtBeginDocument{%
  \hypersetup{
    citecolor=blue,
    linkcolor=blue,   
    urlcolor=blue}}

\usepackage{url}
\usepackage{booktabs}
\usepackage{listings}
\usepackage{paralist} 
\usepackage{wrapfig}
\usepackage{orcidlink}

\usepackage{multirow}
\usepackage{ifpdf}
\usepackage{xspace}
\usepackage{keyval}
\usepackage{color}
\usepackage[font=small,labelfont=bf]{caption}

\definecolor{listinggray}{gray}{0.95}
\definecolor{darkgray}{gray}{0.7}
\definecolor{commentgreen}{rgb}{0, 0.4, 0}
\definecolor{darkblue}{rgb}{0, 0, 0.4}
\definecolor{middleblue}{rgb}{0, 0, 0.7}
\definecolor{darkred}{rgb}{0.4, 0, 0}
\definecolor{brown}{rgb}{0.5, 0.5, 0}

\usepackage[normalem]{ulem}
\makeatletter
\def\cyanuwave{\bgroup \markoverwith{\lower3.5\p@\hbox{\sixly \textcolor{cyan}{\char58}}}\ULon}
\def\reduwave{\bgroup \markoverwith{\lower3.5\p@\hbox{\sixly \textcolor{red}{\char58}}}\ULon}
\def\blueuwave{\bgroup \markoverwith{\lower3.5\p@\hbox{\sixly \textcolor{blue}{\char58}}}\ULon}
\font\sixly=lasy6 %
\makeatother

\newif\ifconceptualmodel
\conceptualmodelfalse

\newif\ifkeepeverything
\keepeverythingfalse

\newif\ifthroughputemulator
\throughputemulatorfalse

\newcommand{\orcid}[1]{\href{https://orcid.org/#1}{\textcolor[HTML]{A6CE39}{\aiOrcid}}}

\newcommand{\pilot}{pilot\xspace}
\newcommand{\pilots}{pilots\xspace}
\newcommand{\pilotedge}{Pilot-Edge\xspace}
\newcommand{\pilotabstraction}{pilot abstraction\xspace}

\newcommand{\up}{\vspace*{-1em}}
\newcommand{\upp}{\vspace*{-0.5em}}

\lstdefinestyle{myListing}{
  frame=single,
  backgroundcolor=\color{listinggray},
  language=C,
  basicstyle=\ttfamily \footnotesize,
  breakautoindent=true,
  breaklines=true
  tabsize=2,
  captionpos=b,
  aboveskip=0em,
  belowskip=-2em,
}

\lstdefinestyle{myPythonListing}{
  frame=single,
  backgroundcolor=\color{listinggray},
  language=Python,
  basicstyle=\ttfamily \scriptsize,
  breakautoindent=true,
  breaklines=true
  tabsize=2,
  captionpos=b,
}

\ifpdf
\DeclareGraphicsExtensions{.pdf, .jpg, .tif}
\else
\DeclareGraphicsExtensions{.ps,  .eps, .jpg}
\fi

\usepackage{listings}
\usepackage{paralist}

\title{Pilot-Edge: Distributed Resource Management Along the Edge-to-Cloud Continuum\upp\upp} 
\author{\\Andre Luckow$^{1,2,3, \orcidlink{0000-0002-1225-4062}}$, Kartik Rattan$^{1}$, Shantenu Jha$^{4,1}$\\
   \footnotesize{\emph{$^{1}$RADICAL, ECE, Rutgers University, Piscataway, NJ 08854, USA}}\\
   \footnotesize{\emph{$^{2}$Ludwig-Maximilian University, Munich, Germany}}\\
   \footnotesize{\emph{$^{3}$Clemson University, South Carolina, USA}}\\
   \footnotesize{\emph{$^{4}$Brookhaven National Laboratory, Upton, NY, USA}\upp\upp\upp\up}
}

\begin{document}

\date{}
\IEEEoverridecommandlockouts
\IEEEpubid{\makebox[\columnwidth]{ xxx~\copyright2020 IEEE \hfill} \hspace{\columnsep}\makebox[\columnwidth]{ }}
\maketitle
\IEEEpubidadjcol

\begin{abstract}
Many science and industry IoT applications necessitate data processing across
the edge-to-cloud continuum to meet performance, security, cost, and privacy
requirements. However, diverse abstractions and infrastructures for managing
resources and tasks across the edge-to-cloud scenario are required. We
propose \pilotedge as a common abstraction for resource management across the
edge-to-cloud continuum. \pilotedge is based on the
\pilotabstraction, which decouples resource and workload management,  and
provides a Function-as-a-Service (FaaS) interface for application-level tasks.
The abstraction allows applications to encapsulate common functions in
high-level tasks that can then be configured and deployed across the continuum.
We characterize \pilotedge on geographically distributed infrastructures using
machine learning workloads (e.\,g., k-means and auto-encoders). Our experiments
demonstrate how \pilotedge manages distributed
resources and allows applications to evaluate task placement based on multiple
factors (e.\,g., model complexities, throughput, and latency).

\end{abstract}


\section{Introduction}\label{sec1}

A growing number of scientific~\cite{nsls,Geernaert} and industrial applications~\cite{doi:10.1002/spe.2816},  require %
the flexible use of resources along the \emph{edge-to-cloud continuum (abbrev.
continuum)}. The coupling of edge and cloud resources enables applications to
address latency, bandwidth, sovereignty, privacy, and security
requirements~\cite{harnessing}. The integration of experimental instruments,
machines, equipment, and other IoT devices, with multiple layers of infrastructures,
 comprising edge, HPC and cloud
infrastructures, is critical to delivering value for these applications.

Machine learning (ML)  methods are essential for
deriving insights from the data produced from these applications. The
combination of growing data volumes and high computational requirements of these
ML applications has accelerated the need for more intelligent use of distributed
computing resources in the
continuum~\cite{DBLP:journals/corr/NishiharaMWTPSL17}. However, these workloads
are highly complex, involving distributed data flows of meta- and raw data, and
the orchestration of inference and training tasks across the continuum. This complexity often results in highly monolithic applications with
tightly coupled application and infrastructure code, limiting the scalability,
reusability, and maintainability of the application.

The complexity, heterogeneity and geographic distribution of IoT, edge, and
cloud infrastructures~\cite{harnessing,nist_fog} make it challenging to design
applications, allocate appropriate resources, and manage workloads.
Particularly, IoT applications are characterized by  heterogeneous tasks,
comprising a mix of real-time tasks for control and steering and long-running
tasks for machine learning training and simulation. Thus, they need to optimize
data and compute placement carefully. Many point and local solutions exist,
which might suffice at small scales, but do not allow for scalable, end-to-end
solutions that permit workload adaptivity and optimization. Complex
application-specific architectures that integrate disperse technological
components lead to unpredictable performance. Thus, it is essential to provide
abstractions that abstract complexity and heterogeneity and enable applications
to adapt to the dynamism induced by  infrastructures, data, and other
sources~\cite{doi:10.1002/cpe.4032}.

This paper introduces the \emph{Pilot-Edge} abstraction and framework.
\pilotedge is motivated by improved
edge-to-cloud application development, deployment, and management. It provides 
a \emph{Function as a Service (FaaS)} interface which abstracts resources 
from the application.
\pilotedge allows applications to decompose workloads into 
tasks, and deploy them  across the continuum. 
\pilotedge orchestrates 
tasks generated from the function code, handling placement and data movements
transparently, considering application-defined preferences (e.\,g.,  
data dependencies and preferred placements). 
\pilotedge relies on the \pilotabstraction for distributed
resource management and unified access to resources across all layers. We
envision \pilotedge as providing the
hierarchical but continuous resource management fabric for edge-to-cloud infrastructures, enabling many
increasingly complex applications comprising heterogeneous multi-task workloads,
and requiring diverse resource capabilities.

\pilotedge was designed based on an analysis of different IoT application
scenarios (e.\,g., earth sciences, light source
science~\cite{edge_emulation_2021}). It enables the effective handling of
heterogeneous and dynamic workloads arising in IoT environments (e.\,g., 
seasonal peak loads, failures and other external events). 
\pilotedge allows applications to respond to dynamism,
e.\,g., external events, load peaks, and resource failures, by updating their tasks'
payload or acquiring additional resources. We characterize and demonstrate
the capabilities of \pilotedge using extensive end-to-end experiments on
geographically distributed infrastructure, particularly XSEDE
(US)~\cite{jetstream} and LRZ (Germany), and ML workloads, e.\,g.,
auto-encoders.

This paper is structured as follows: In section~\ref{sec:pilotedge}, we present
Pilot-Edge and provide an extensive evaluation of the framework in section~\ref{sec:experiments}.
Section~\ref{sec:related} discusses related work.

\section{Pilot-Edge: Abstraction and Framework}
\label{sec:pilotedge}

Managing the edge-to-cloud continuum's complexity and dynamism requires a 
sophisticated framework that aids in managing resources and
workloads.  \emph{\pilotedge} aims to simplify the development of edge-to-cloud
applications by providing a high-level abstraction for developing, deploying,
and managing computation and data across multiple layers of distributed
infrastructure. After discussing previous work on the \pilotabstraction in section~\ref{subsec:pilotabstraction}, we present \pilotedge's architecture and abstraction in sections~\ref{sec:architecture} and \ref{sec:abstraction}.

\subsection{Previous Work: Pilot-Abstraction}
\label{subsec:pilotabstraction}

\pilotedge is based on the \pilotabstraction, an abstraction for distributed
resource management~\cite{pstar12}. The \pilotabstraction is based on the
observation that using a placeholder job to allocate a resource container is a
re-occurring pattern used by many applications. The \pilotabstraction decouples
resource and workload management and supports manifold workloads, particularly
workloads that require task parallelism on HPC and clouds. The term \pilot
refers to a placeholder job in a queuing system that allocates resources on
which the application can execute tasks. A pilot generally refers to a dedicated
resource set that an application owns, e.\,g., a virtual machine, a job
partition (HPC), or a Lambda function~\cite{luckow2019performance}.

While the \pilotabstraction was designed for HPC, we extended it for
data-intensive and streaming applications, which similarly exploit data
parallelism.  Pilot-Data added support for data management in conjunction with
\pilots. Further, we integrated frameworks for data
processing~\cite{2016arXiv160200345L,pilot-data-jpdc-2014}, such as Spark and Dask~\cite{dask}, and
streaming~\cite{pilot-streaming}, such as Kafka~\cite{kreps2011kafka}.
Pilot-Streaming also allows the event-driven execution of tasks on-demand, 
e.\,g., responding  to data arrival events.

While the \pilotabstraction is well suited for bridging heterogeneous
infrastructures across the edge-to-cloud continuum and
administrative domains, the current implementation has several limitations: (i)
the provided abstraction is low-level, requiring applications to manage resources
and wrap their workload into tasks, and (ii) the implementation is optimized for
data-center-based infrastructure and workloads.

\subsection{Architecture}
\label{sec:architecture}

\begin{figure}[t]
	\centering
	\includegraphics[width=0.49\textwidth]{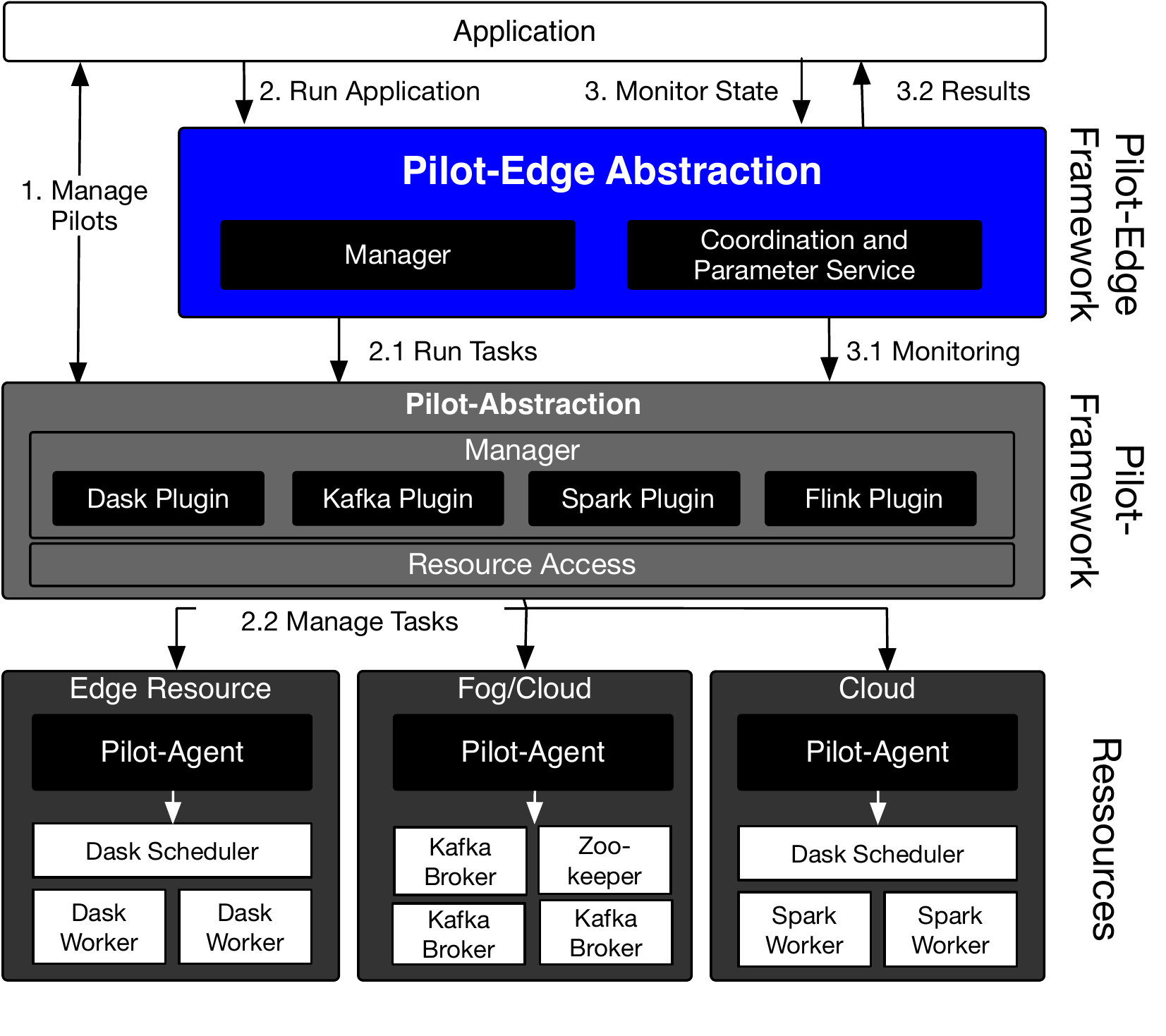}
  \caption{\textbf{\pilotedge Architecture and Interactions:} \pilotedge
  comprises the \pilotedge framework and \pilot framework. Applications
  acquire edge-to-cloud resources using the \pilot framework in step 1. In
  step 2 applications configure \pilotedge using the resources acquired and
  submit their workload to the framework. Comprehensive monitoring services
  are provided (step 3). 
  \label{fig:pilotedge_architecture}\up\up}
\end{figure}

\pilotedge extends Pilot-Streaming~\cite{pilot-streaming} and supports various
resource types via a plugin-based architecture, e.\,g., HPC and cloud clusters
(such as OpenStack, AWS), smaller IoT devices (via SSH). Further, \pilotedge
extensively utilizes message brokering based on Kafka to manage edge-to-cloud
streaming topologies. Brokering concerns are also encapsulated using a plugin
mechanism. Support for further brokering framework, e.\,g., MQTT for
low-performance and low-power environments, can easily be added.

Figure~\ref{fig:pilotedge_architecture} illustrates the overall architecture
(blue components extended in this work). A typical application 
comprises  three stages: (step 1) allocating resources using the \pilotabstraction,
(step 2) running a distributed edge-to-cloud application, and (step 3) monitoring
applications.

As edge-to-cloud applications typically rely on highly-specialized processing
pipelines and resources, we currently require the manual allocation of resources
via the \pilotabstraction~\cite{pstar12} (step 1 in
Fig.~\ref{fig:pilotedge_architecture}).  The \pilotabstraction provides a common interface to allocate arbitrary resources, e.\,g., a
RasPi, virtual machines located on the edge and cloud, serverless cloud
functions, HPC machines. Thus, depending on the continuum layer, a pilot can
represent different types of resources. Further, the \pilotabstraction can
manage brokering and data processing frameworks, e.\,g., Kafka and Dask. In
summary, the \pilotabstraction encapsulates much of the complexity of
distributed resources. The created pilots are then used to initialize
the application (step 2 in Fig.~\ref{fig:pilotedge_architecture}).

After submitting the application, \pilotedge translates and packages the
user-defined functions into tasks to be executed on the edge, cloud, or HPC
pilots (step 2). Further, it provides a central coordination and parameter service 
to share state, e.\,g., for data or machine learning models, 
across the continuum.  \pilotedge automatically handles
task placements, i.\,e., the binding of a task to a pilot (step 2.1).

The tasks are executed using a managed Dask~\cite{dask} cluster on the specified location
(step 2.2).  The input data is passed as a parameter to each function; the output is captured with a
return parameter. Further information on the resource topology and shared state
are via a \texttt{context} object. A unique job identifier
ensures that progress and errors can be consistently tracked across all
components. The framework also manages the data movements using a pilot-managed Kafka broker and  an automatically created Kafka topic. Further, it provides a
Redis-based parameter server for sharing model weights across the continuum. 

\subsection{Pilot-Edge API}
\label{sec:abstraction}

\pilotedge exposes a \emph{Function-as-a-Service (FaaS) API}, 
that abstracts details about individual resources, allowing the application to 
focus on application logic and not infrastructure. While the framework 
is suited to support arbitrary IoT edge applications, we mainly focus on data
and machine learning applications.

\begin{lstlisting}[basicstyle=\scriptsize, caption=Pilot-Edge FaaS API, label=lstedgetocloud_api] 
  def produce_edge(context)
  
  def process_edge(context: dict = None, data=None)
  
  def process_cloud(context: dict = None, data=None)
\end{lstlisting}

Listing~\ref{lstedgetocloud_api} illustrates the API of the
\pilotedge-abstraction.  The API is application-centric and lets developers
focus on expressing important application tasks, e.\,g., sensing and
inference, and on selected trade-offs, such as task localities. The API comprises three
functions: (i) for managing sensing and data generation on the edge, (ii) for
edge processing, and (iii) for cloud processing. While each task must be defined
as a Python function, it is also possible to access native capabilities, e.\,g.,
by integrating native code for accessing low-level sensors on the edge. The API allows the re-use of functions across the continuum while retaining flexibility and customizability.

\begin{lstlisting}[basicstyle=\scriptsize, caption=\pilotedge API: Instantiation of an Application, label=lstedgetocloud] 
pilot.EdgeToCloudPipeline  (
  pilot_cloud_processing=pilot_job_cloud_processing, 
  pilot_cloud_broker=pilot_job_cloud_broker,
  pilot_edge=pilot_job_edge,
  produce_function_handler=produce_block_edge,
  process_edge_function_handler=process_block_edge,
  process_cloud_function_handler=process_block_cloud,
  function_context=context, 
  ...
  ).run()
\end{lstlisting}

Listing~\ref{lstedgetocloud} shows how an edge-to-cloud application is
instantiated. In addition, to passing the function references to the data
generation and processing functions, a references to the edge and cloud pilot is
required. The framework then handles the dataflow between the instantiations of
the defined functions in these pilots using Kafka.

\upp
\subsection{Discussion}

\pilotedge provides a blueprint for applications and supports common patterns,
e.\,g., integrating sensing tasks, i.\,e., tasks that capture environmental
changes using sensors, and other types of processing, e.\,g., pre-processing and
machine learning inference. For example, commonly, the API's data source function  
(\texttt{produce\_edge} in Listing~\ref{lstedgetocloud_api}) is used either to deploy data
collection code, e.\,g., code for reading out a sensor or a data generator. The
edge and cloud functions are used for processing. For example, the edge function
frequently serves for data pre-aggregation, outlier detection, and data
compression to ensure that the amount of data movement is minimal. The cloud
functions are often used for more complex analytics, training, and modeling tasks.

By nature, edge-to-cloud applications are subject to different dynamism and
variability induced by data sources, infrastructures, and applications. If
supported by the resource, the allocated resources can be adapted, i.\,e.,
expanded and scaled-down, dynamically at runtime, e.\,g., if a bottleneck arises
due to increased data rates or in response to an application event (e.\,g., the
discovery of a significant data pattern). The processing functions can be
programmatically replaced at runtime (without the need to allocate a new pilot),
allowing, e.\,g., the exchanging low vs. high fidelity models.

\begin{figure}[t]
  \upp
	\centering
	\includegraphics[width=0.5\textwidth]{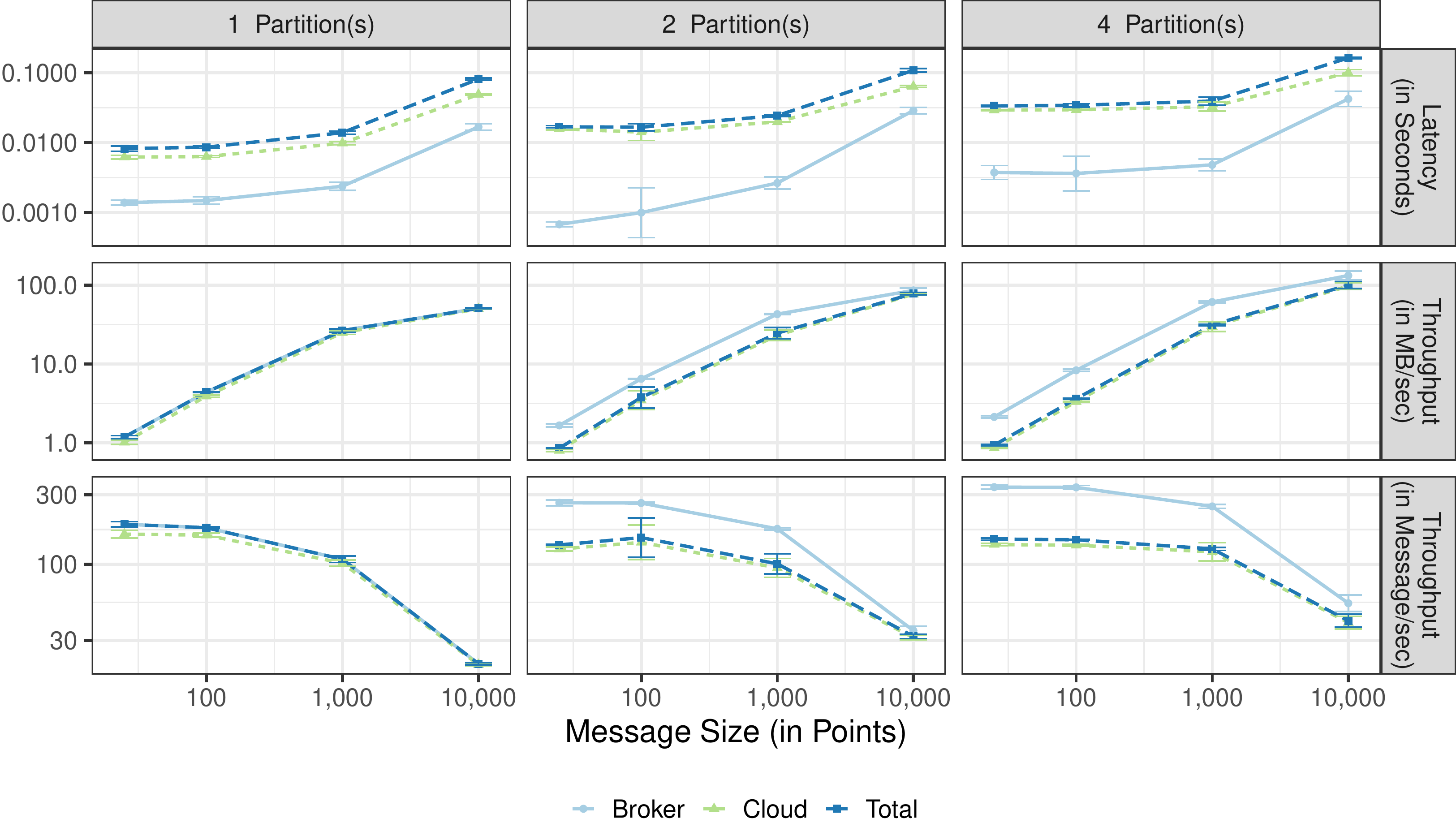}
	\caption{\textbf{Throughput and Latencies by Message Size and Partitions:} The system's total throughput increases with the number of edge devices and partitions; every edge device is assigned a dedicated partition. In the four partition scenario, the processing system becomes the bottleneck determining the overall throughput.\label{fig:throughput}\up\up}
\end{figure}

\section{Experiments}
\label{sec:experiments}

This section conducts a performance characterization of different machine learning workloads using \pilotedge.  For our experiments, we use the Leibniz Supercomputing Center (LRZ) und XSEDE Jetstream clouds and different VM types:  4\,core/18\,GB (medium), 10 cores/44\,GB (large) (LRZ) and 6\,cores/16\,GB (medium) (Jetstream). Synthetic data is generated using the Mini-App data generator~\cite{luckow2019performance}.

\subsubsection{Baseline Performance}

We investigate the throughput and latency with the edge data source, broker, and processing components deployed on the LRZ cloud. The edge devices are simulated with a Dask task, allocating one core and about 4\,GB of memory, comparable to a current Raspberry Pi.  We use one partition per edge device for simplicity and keep the ratio of partitions constant between Kafka and Dask. We use message sizes of 25 to 10,000 points with 32 features each. Every point has a serialized size of 8\,Bytes, i.\,e.,  message sizes are 7\,KB to 2.6\,MB. We send 512\,messages per run and repeat each experiment at least three times.

Figure~\ref{fig:throughput} illustrates the baseline throughput and latencies. The framework captures and links comprehensive metrics across all involved components, particularly the edge data generator, broker, and cloud processing services (for clarity, data for edge is not displayed). This data allows the easy identification of bottlenecks. For example, for four partitions, it is apparent that the Kafka broker can process more data than the consuming processing tasks in the cloud.

\subsubsection{Machine Learning Models and Geographic Distribution}

We continue to evaluate three machine learning models for outlier detection. We primarily use the cloud-centric deployment pattern (see Figure~1 in~\cite{edge_emulation_2021}), i.\,e., we  deploy the data generator on the edge and the processing tasks, which include pre-processing, training and inference, on the cloud.  We evaluate three machine learning models: the auto-encoder, isolation forests, and k-means (25 clusters as previously). In all cases, the model is updated based on the incoming data; model updates are managed via the parameter service. We use the large VM on LRZ for all processing tasks (10 core/44\,GB).

Isolation forests~\cite{4781136} are an ensemble technique where each task partitions the dataset randomly into trees. An outlier is defined by the number of steps required to isolate a data point; the fewer steps required, the more likely a point is an outlier. We use the PyOD~\cite{zhao2019pyod} implementation and a default of 100 ensemble tasks. Auto-encoders~\cite{10.5555/3086742} are unsupervised models that rely on a deep neural network to learn a data representation. For outlier detection, the reconstruction error is used to determine whether a data point is anomalous. We use the Keras-based auto-encoder implementation of PyOD with four hidden layers with a size of [64, 32, 32, 64], and thus, a total number of 11,552 parameters.

\begin{figure}[t]
	\centering
	\includegraphics[width=0.5\textwidth]{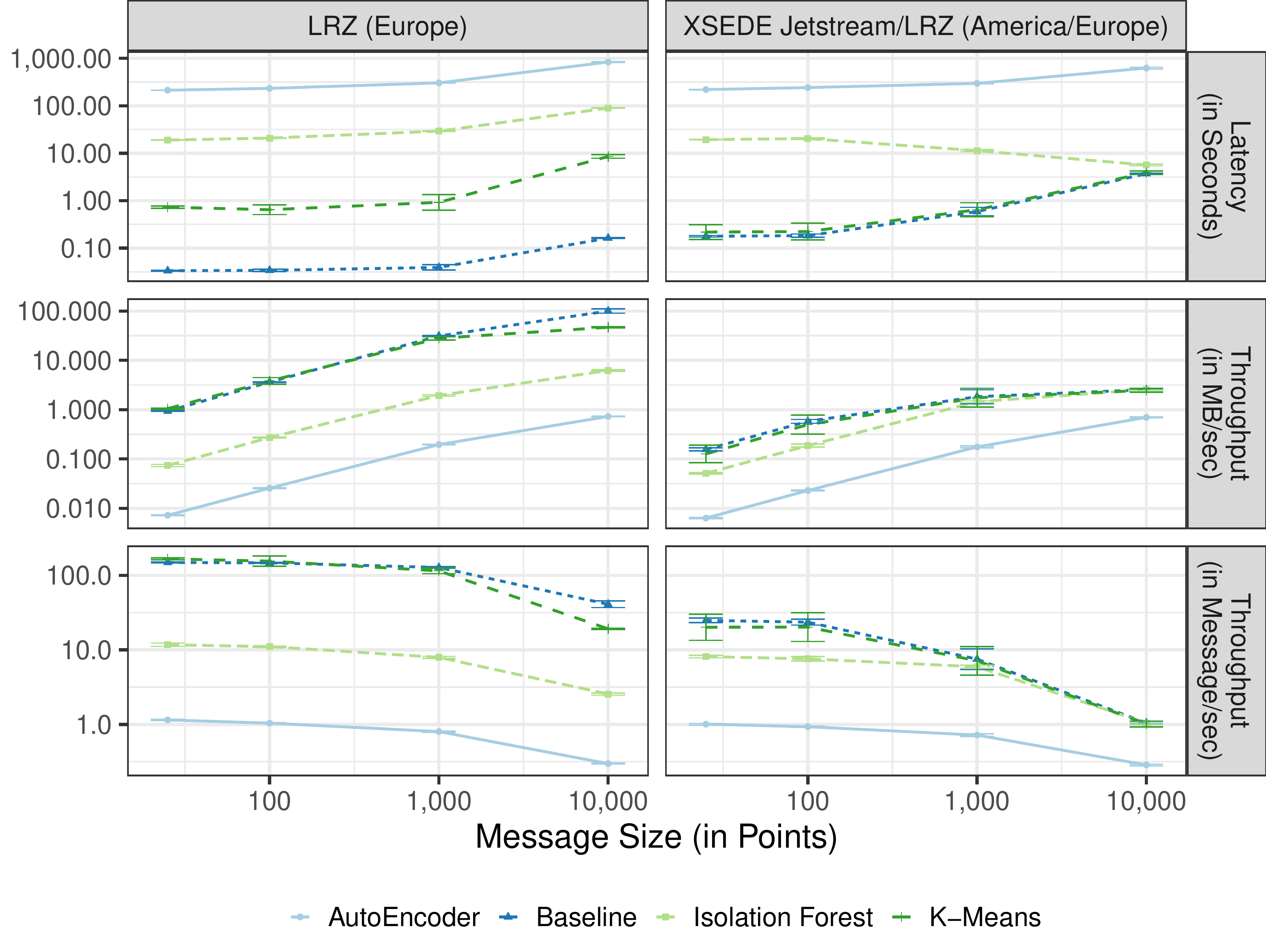}
	\caption{\textbf{Throughput and Latency by Model Type, Message Size, and Geographical Distribution:} The model complexity significantly impacts all metrics. K-means outperforms isolation forests and  auto-encoders, which shows the worst performance and is not well suited for environments with limited resources.\label{fig:throughput_latency_algorithms}\up\up}
\end{figure}

Figure~\ref{fig:throughput_latency_algorithms} illustrates that as the computational complexity increases, the performance degrades significantly compared to the baseline case.  Isolation forests achieve a significantly worse performance than k-means for both  latency and throughput. Auto-encoders required careful tuning of the system; we had to adjust the memory and garbage collection.  Due to their high resource demands, they are not suitable for streaming and require additional resources, e.\,g., GPUs. Alternatively, an edge or hybrid deployment would be an option.

Further, we investigate the geographic distribution and place the data source on Jetstream/XSEDE (US) and processing stages (i.\,e., pre-processing, training, and inference) on the LRZ cloud (Europe).  The latency between both locations varied between 140 and 160 msec; bandwidth fluctuated between 60 to 100 MBits/sec (iPerf measurement). We use four partitions for this experiment. As expected, the overall throughput for the baseline and k-means scenarios is limited by intercontinental data transfer. Both scenarios would benefit from a hybrid edge-to-cloud deployment, e.\,g., by adding a data compression step before the data transfer. The results also show that the network is not the bottleneck for the compute-intensive models, i.\,e., auto-encoder and isolation forests.

\section{Related Work}
\label{sec:related}

Apache Edgent~\cite{edgent} is an edge platform designed to integrate IoT devices and brokering systems, e.\,g., Kafka.  Edgent is narrowly focused on the edge device and broker communication and does not consider resource management across the continuum holistically. Similarly, SpanEdge~\cite{7774704} is a stream processing system based on Apache Storm~\cite{10.1145/2588555.2595641} that allows applications to run certain parts of a topology, a Storm processing pipeline, close to the data source. SpanEdge dependence on Storm limits its applicability in the heterogeneous continuum. The programming model lacks many aspects required for machine-learning-based applications, e.\,g., the integration with modern Python frameworks like Tensorflow and Dask.

Further, various Kubernetes-based frameworks emerged, e.\,g., MicroK8s and KubeEdge~\cite{kubeedge}. KubeEdge extends  containerized application orchestration and device management to the edge. 
While Kubernetes is cloud-agnostic and provides some interoperability, it is also highly complex and designed for a stable cloud environment. Data-related concerns, such as data movements, are not transparently handled and need to be implemented on the application-level.

Different public cloud providers offer edge extensions for their serverless
FaaS runtime.  For example, Lambda Edge~\cite{lambda_edge} enables the
execution of Lambda function in Greengrass IoT runtimes. A similar offering
exists on Azure with IoT Edge~\cite{azure_iot_edge}.  While FaaS is easy to
use and benefits from the automatic resource management and scaling of clouds,
these benefits do not apply necessarily to edge devices, subject to
significant resource constraints. Further, several research frameworks that
explore the usage of FaaS along the edge-to-cloud continuum emerged, e.\,g.,
CSPOT~\cite{10.1145/3318216.3363314} and funcX~\cite{10.1145/3369583.3392683}.

While these related frameworks offer similar abstractions, \pilotedge differs in different
aspects: (i) \pilotedge supports highly heterogeneous workloads and infrastructures, bringing together distributed resources and capabilities from different providers. 
(ii) By decoupling resource management and application-level
scheduling, applications can better respond to dynamic changes in the
environment. 
(iii) \pilotedge provides more flexible mechanisms to handle data
and models across the continuum, e.\,g., by integrating brokering services for
data streaming and coordination services for sharing machine learning models.

\section{Conclusion and Future Work}

We presented \pilotedge, an abstraction for supporting data and ML applications
in the edge-to-cloud continuum addressing the following challenges: (i)
\emph{Heterogeneity:}  The edge-to-cloud continuum  is highly diverse, comprising many different types of hardware and software components that need to be unified and integrated, 
(ii) \emph{Dynamism} in distributed,
geographically disperse environments often constrains application leading to
unacceptable and unpredictable performance. The ability to respond  at
runtime, e.\,g., by auto-scaling resources, is crucial, and (iii)
\emph{Performance} in distributed, heterogeneous environments can be highly
unpredictable depending on shared resources,  system loads, and data.

Pilot-Edge was designed based on an analysis of different applications and
provides an easy-to-use FaaS API that simplifies application development,
allowing developers to focus on application logic and application-level resource
management. It supports common deployment modalities, e.\,g., more cloud-centric
or edge-centric scenarios. Tasks can easily be moved to different parts of the
continuum at runtime. Particularly, it supports common data collection, model
training, and deployment patterns of ML-driven IoT applications.

Our experiments investigated various trade-offs, e.\,g., the impact of model
complexity on the overall throughput. For example, k-means can achieve five
times the throughput of isolation forests for large message sizes (10,000
points). Further, auto-encoders proved unsuitable for the investigated resource
configurations due to their high computational demands. These insights provide
valuable input for system design and deployment, allowing an optimal resource
layout.

In the future, we will continue to extend \pilotedge and simplify the usage.
For example, we will generalize the abstraction to arbitrary architectures and
topologies of resources \textemdash{} currently, it is limited to two layers:
edge and cloud. 
We envision \pilotedge as the basis for a distributed workload  management
system that can select, acquire and dynamically scale resources across the
continuum at runtime based on the application's objectives. To further enhance
our understanding of the continuum, we will explore novel edge-to-cloud
scenarios, e.\,g., federated learning, and investigate further scheduling and 
approaches, e.\,g., energy consumption.

\vspace{2mm}
\subsubsection*{Acknowledgments} 

\footnotesize 
Computational resources were provided by NSF XRAC award TG-MCB090174 and the
Leibniz Supercomputing Center (LRZ). This work was supported by a DOE SBIR
Award DE-SC0017047 to Optimal Solutions Inc.

\bibliographystyle{unsrt}
\bibliography{pilotedge,radical_publications,streaming}

\end{document}